\documentclass[11pt]{article}
\usepackage{graphicx}
\usepackage{latexsym}
\usepackage{amssymb}
\begin{document}

\def \d {{\rm d}}
\def \U {{\cal U}}
\def \V {{\cal V}}
\def \H {{\cal H}}
\def \M {{\cal M}}
\def \e {\varepsilon}
\def \E {{\bf e}}
\def \m  {{\bf m}}
\def \bl {{\bf l}}
\def \k  {{\bf k}}
\def \uu {{\bf u}}
\def \R  {{\cal R}}
\def \I  {{\cal I}}
\def \A  {{\cal A}}
\def \C  {{\cal C}}
\newcommand{\be}{\begin{equation}}
\newcommand{\ee}{\end{equation}}
\newcommand{\beqn}{\begin{eqnarray}}
\newcommand{\eeqn}{\end{eqnarray}}
\newcommand{\pa}{\partial}
\newcommand{\pp}{{\it pp\,}-}

\title{Impulsive waves in electrovac direct product  spacetimes with $\Lambda$}

\addvspace{1cm}

\author{
Marcello Ortaggio\thanks{E--mail: {\tt ortaggio@science.unitn.it}} \\
Dipartimento di Fisica, Universit\`a degli Studi di Trento,  \\
and INFN, Gruppo Collegato di Trento, 38050 Povo (Trento), Italy. 
\\ 
and
\\ 
Ji\v{r}\'{\i} Podolsk\'y\thanks{E--mail: {\tt podolsky@mbox.troja.mff.cuni.cz}}\\
 Institute of Theoretical Physics, Charles University in Prague,\\
V Hole\v{s}ovi\v{c}k\'ach 2, 18000 Prague 8, Czech Republic. 
}

\date{\today}

\maketitle
\begin{abstract}
A complete family of non-expanding impulsive waves in spacetimes which are the direct product of two  2-spaces of constant curvature is presented. In addition to previously investigated impulses in Minkowski, (anti-)Nariai and Bertotti--Robinson universes, a new explicit class of impulsive waves which propagate in the exceptional electrovac Pleba\'nski--Hacyan spacetimes  with a cosmological constant $\Lambda$ is constructed. 
In particular, pure gravitational waves generated by null particles with an arbitrary multipole structure are described. The metrics are impulsive members of a more general family of the Kundt spacetimes of type~$II$.  The well-known \pp waves are recovered for $\Lambda=0$.


\end{abstract}


\section{Introduction}

A class of exact solutions which represent non-expanding impulsive waves in the Nariai universe has recently been introduced in \cite{Ortaggio01}. The geometrical and physical properties of the impulse have also been investigated in detail. In addition, it has been suggested in \cite{Ortaggio01} how to extend the construction to other well-known direct product spacetimes, namely the anti-Nariai and Bertotti--Robinson universes. It is the purpose of the present paper to perform such extension explicitly and to demonstrate that, in fact, it can be generalized to {\em all} spacetimes which are the direct product of two 2-spaces of constant curvature. After presenting a general family of exact non-expanding impulsive waves propagating in such an arbitrary direct product background (this section), we shall concentrate on the two possibilities for which \emph{one and only one} of the 2-spaces has a vanishing curvature (section 2). These are two of the three exceptional Pleba\'nski--Hacyan spacetimes \cite{PlebHac79} which were not considered previously as backgrounds for impulses. Finally, we briefly discuss further generalizations to the third Pleba\'nski--Hacyan spacetime (which is not a direct product) and to finite sandwich waves (section 3).

We consider here the class of spacetimes obtained by constraining a 6-dimensional impulsive \pp wave
\be
  \d s^2= 2\,\d U\d V+\epsilon_1\d{Z_2}^2+\d{Z_3}^2+\d{Z_4}^2+\epsilon_2\d{Z_5}^2
    -H(Z_2,Z_3,Z_4,Z_5)\,\delta(U)\,\d U^2  \ ,
\ee
to the 4-submanifold given by
\beqn
    2\epsilon_1UV+{Z_2}^2=a^2 \ , \qquad \epsilon_2({Z_3}^2+{Z_4}^2)+{Z_5}^2=b^2\ ,\label{cons}
\eeqn
where $a, b$ are positive constants, and $\ \epsilon_1, \epsilon_2=0,+1,-1\,$. Natural coordinates are introduced by the parameterization  
\beqn
 & & U=\frac{u}{\Omega}\  , \qquad V=\frac{v}{\Omega}\  , 
 \qquad Z_2=a\frac{1-\frac{1}{2}\epsilon_1a^{-2}uv}{\Omega}\  , \nonumber  
 \label{nullcoord} \\
 & & Z_3=\frac{\zeta+\bar{\zeta}}{\sqrt2\,\Sigma}\  , \quad Z_4=-i\frac{\zeta-\bar{\zeta}}{\sqrt2\,\Sigma}\ , \quad Z_5=b\frac{1-\textstyle{\frac{1}{2}}\epsilon_2b^{-2}\zeta\bar{\zeta}}{\Sigma}\  ,
\eeqn
where
\begin{equation}
\Omega=1+\textstyle{\frac{1}{2}}\epsilon_1a^{-2}uv\ , \qquad 
\Sigma=1+\textstyle{\frac{1}{2}}\epsilon_2b^{-2}\zeta\bar{\zeta}\ ,
\end{equation}
in which the metric takes the form
\beqn
\d s^2=\frac{2\,\d\zeta\d\bar{\zeta}}{(1+\frac{1}{2}\epsilon_2b^{-2}\zeta\bar{\zeta})^2}
+\frac{2\,\d u\d v - H(\zeta,\bar{\zeta})\,\delta(u)\,\d u^2}{(1+\frac{1}{2}\epsilon_1a^{-2}uv)^2} \ .
\label{metric}
\eeqn
Using the  null tetrad ${\bf m}=\Sigma\,\pa_{\bar{\zeta}}$, 
${\bf l}=-\Omega\,[\pa_u+\frac{1}{2}H\delta(u)\pa_v\,]$, ${\bf k}=\Omega\,\pa_v\,$, we obtain the only non-vanishing Weyl and Ricci scalars
\beqn
&& \Psi_2  =  -{\textstyle\frac{1}{6}}\left(\frac{\epsilon_1}{a^2}+\frac{\epsilon_2}{b^2}\right)\ , \qquad  
\Phi_{11}={\textstyle\frac{1}{4}}\left(-\frac{\epsilon_1}{a^2}+\frac{\epsilon_2}{b^2}\right) \ ,\nonumber  \\
&& \Psi_4  = {\textstyle\frac{1}{2}}\left(\Sigma^2 H_\zeta\right)_\zeta\delta(u) \ , \hskip7.2mm
\Phi_{22}  = {\textstyle\frac{1}{2}}\Sigma^2H_{\zeta\bar{\zeta}}\,\delta(u)\ , \label{R} \\  
&& R  =  2\left(\frac{\epsilon_1}{a^2}+\frac{\epsilon_2}{b^2}\right)\ . \nonumber
\eeqn
It is now obvious that the above solutions represent exact impulsive waves localized on the null 3-submanifold  $U=0=u$.  The impulse, which consists of gravitational and/or matter components, propagates in various possible backgrounds. These correspond to the metric (\ref{metric}) with $H=0$ and, obviously, represent all spacetimes which are a direct product of two 2-spaces of arbitrary constant curvature $K_1=\epsilon_1a^{-2}$ and $K_2=\epsilon_2b^{-2}$, respectively. All the physically reasonable backgrounds are summarized in the following table~1 and schematically in figure~1.
Recall that these well-known universes [2--7] 
 are either vacuum or contain a uniform electromagnetic field, possibly with a cosmological constant $\Lambda$.
\begin{table}[htbp]
 \[
  \begin{array}{|c|c||c|c|} \hline
\ \epsilon_1\  & \ \epsilon_2\  & \mbox{ geometry } & \mbox{ background universe } \\ \hline\hline
   0 & 0 & \mbox{ }\mathbb{R}^2_1\times\mathbb{R}^2 & \mbox{ Minkowski} \\ \hline      
   +1 & +1 & \mbox{ dS}_2\times\mathbb{S}^2 & \mbox{ Nariai} \\ \hline
   -1 & -1 & \mbox{ AdS}_2\times{\mathbb H}^{\,2} & \mbox{ anti-Nariai} \\ \hline
   -1 & +1 & \mbox{ AdS}_2\times\mathbb{S}^2 &   \mbox{ Bertotti--Robinson} \\ \hline
   0 & +1& \mbox{ }\mathbb{R}^2_1\times\mathbb{S}^2 & \mbox{ Pleba\'nski--Hacyan ($\Lambda>0$)} \\ \hline      
   -1 & 0 & \mbox{ AdS}_2\times\mathbb{R}^2 & \mbox{ Pleba\'nski--Hacyan ($\Lambda<0$)} \\ \hline
  \end{array}
 \]
 \caption{Possible background spacetimes which are the direct product of two constant-curvature 2-spaces. The remaining three  choices of $\epsilon_1$, $\epsilon_2$ are unphysical since the energy density $\Phi_{11}$ would be negative. }
 \label{}
\end{table}
\begin{figure}
 \centering
 \includegraphics[width=0.6\textwidth]{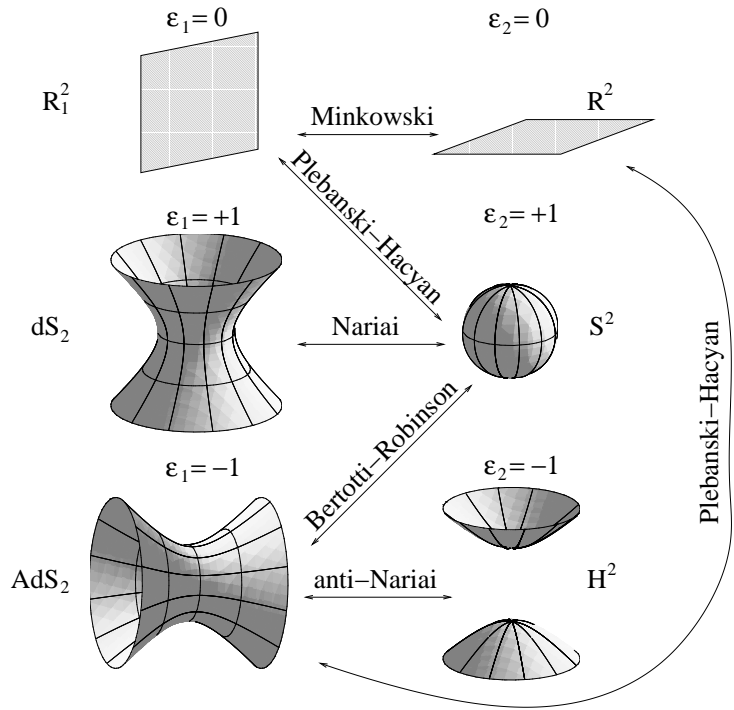}
 \caption{The possible spacetimes containing non-expanding impulsive waves (\ref{metric}). The parameter $\epsilon_1$ determines the conformal structure of the background, whereas $\epsilon_2$ gives the geometry of the impulse. (The figure is inspired by \cite{Bertotti59}.)}
\end{figure}

By considering a non-trivial function $H$, impulsive waves are introduced into the above universes. In the simplest case $\>\epsilon_1=0=\epsilon_2\,$, the metric (\ref{metric}) reduces to the standard form of famous impulsive \pp waves with planar wavefronts propagating in the Minkowski space \cite{penrosestructure} (see, e.g., \cite{Podolsky02} for more references). For non-vanishing $\,\epsilon_1\,$ and/or $\,\epsilon_2\,$ one obtains impulses in the curved backgrounds summarized in table~1, i.e. in all spacetimes which are the direct product of two 2-spaces of constant curvature.

\section{Impulsive waves in the Pleba\'nski--Hacyan spacetimes}

First, let us observe from the general relation $R=4\Lambda-8\pi T$ and the expression for $R$ given in (\ref{R})  that the trace $T$ of the energy-momentum tensor has the \emph{same} constant value everywhere. Moreover, for  $u\not=0$ the matter field of the background satisfies the Maxwell equations so that $T=0$. Therefore, $R\equiv 4\Lambda$  and $\Psi_2 \equiv -\frac{1}{3}\Lambda$ \emph{everywhere}, including on the impulse localized on $u=0$.
The expression for $R$ in (\ref{R})  can thus be written as
\be
\frac{\epsilon_1}{a^2}+\frac{\epsilon_2}{b^2}=2\Lambda\ .\label{cond}
\ee
The two cases $\>\epsilon_1=0$, $\epsilon_2=+1\,$ and $\>\epsilon_1=-1$, $\epsilon_2=0\,$, which we wish to investigate here, thus necessarily require $\Lambda>0$ and $\Lambda<0$, respectively. Moreover, it follows from (\ref{R}) that $\Phi_{11}=\pm\frac{1}{2}\Lambda>0$ so that the curvature scalars satisfy $2\Phi_{11}\pm 3\Psi_2=0$. 
The background spacetimes are thus exactly two of the ``exceptional electrovac type $D$ metrics with cosmological constant'' investigated by Pleba\'nski and Hacyan \cite{PlebHac79}.\footnote{All type $D$ solutions of the Einstein-Maxwell equations in the presence of $\Lambda$ were investigated by Pleba\'nski \cite{Plebanski79} under the assumptions that both double principal null directions are non-expanding, non-twisting and aligned along the real eigenvectors of the (non-null) electromagnetic field. When $2\Phi_{11}\pm 3\Psi_2\not =0$, the Bianchi identities imply that the principal null directions are also geodesic and shear-free. The exceptional cases  $2\Phi_{11}\pm 3\Psi_2=0$ were analyzed in detail in \cite{PlebHac79}.}

\subsubsection*{The $\Lambda>0$ case}

For $\>\epsilon_1=0$, $\epsilon_2=+1\,$ it follows from (\ref{cond}) that
$\frac{1}{2}b^{-2}=\Lambda$. The metric (\ref{metric}) thus reduces to
\be
 \d s^2={2\,\d\zeta\d\bar\zeta\over \left(1+\Lambda\zeta\bar\zeta\right)^2}+2\,\d u\d v-H(\zeta,\bar\zeta)\,\delta(u)\,\d u^2 \ ,
 \label{four+}
\ee
and the Weyl and Ricci scalars which represent radiation become
$\Psi_4=\textstyle{\frac{1}{2}}[\left(1+\Lambda\zeta\bar\zeta\right)^2H_\zeta]_\zeta\delta(u)$,
$\Phi_{22}=\textstyle{\frac{1}{2}}(1+\Lambda\zeta\bar\zeta)^2H_{\zeta\bar{\zeta}}\,\delta(u)$.
On the wave front $u=0$, the spacetime (\ref{four+}) is in general of the Petrov type $II$ and represents an impulsive gravitational wave plus an impulse of pure radiation. These propagate in the Pleba\'nski--Hacyan universe which is the direct product $\mathbb{R}^2_1\times\mathbb{S}^2$ of a 2-Minkowski space with a 2-sphere, thus admitting a six-dimensional group of isometries $ISO(1,1)\times SO(3)$.  The impulse describes the history of a {\em non-expanding 2-sphere} of a constant area $2\pi/\Lambda$. 
Notice also that $\pa_v$ is a Killing vector of (\ref{four+}) for an arbitrary $H$.

In particular, when $H$ satisfies $\Phi_{22}=0$ there is no impulsive pure radiation, and the metric (\ref{four+}) thus represents a purely gravitational impulsive wave propagating in the electrovac background. 
Interestingly, this equation is exactly the same as that discussed in the  context of the Nariai universe.
It has a simple general solution $H(\zeta,\bar{\zeta})=f(\zeta)+\bar{f}(\bar{\zeta})$ (where $f(\zeta)$ is an arbitrary analytic function of $\zeta$) which, unless it is a constant, necessarily contains singularities. These are localized on the spherical wavefront and, following \cite{GriPod97,PodGri98cqg,Ortaggio01}, can naturally be considered as null point sources of the impulsive gravitational wave. In order
to achieve such a physical interpretation, it is convenient to use the coordinates $(z,\phi)$ on the sphere defined by $\zeta=\Lambda^{-1/2}\sqrt{(1-z)/(1+z)}\,e^{i\phi}$. The general solution can thus be rewritten as (cf. \cite{Ortaggio01})
\be
 H(z,\phi)=a_0+{\textstyle\frac{1}{2}}b_0\ln\frac{1+z}{1-z}+\sum_{m=1}^\infty \Big( b_mF_m(z)+ b_{-m}F_{-m}(z)\Big)\cos[m(\phi-\phi_m)] \ ,
 \label{purewaves+}
\ee
where $a_0$, $b_0$, $b_{\pm m}$ and $\phi_m$ are arbitrary constants, and 
$ F_{\pm m}(z)\equiv (1-z^2)^{m/2}\frac{\d^m}{\d z^m}\ln(1\mp z)^{1/2} $.
The constant term $a_0$ can be removed by the discontinuous transformation $v\rightarrow v+\frac{1}{2}a_0\Theta(u)$. Non-trivial solutions (\ref{purewaves+}) thus contain at least one singularity at $z=1$ or $z=-1$, i.e. at one of the poles of the spherical wave surface. 
If we define a source distribution $J(z,\phi)$ by $\Phi_{22}\equiv\frac{1}{2}\Lambda\,J(z,\phi)\,\delta(u)$, and substitute (\ref{purewaves+}) into the expression (\ref{R}) for $\Phi_{22}$, we obtain $J(z,\phi)=b_0J_0(z)+\sum_{m=1}^\infty[b_mJ_m(z,\phi)+b_{-m}J_{-m}(z,\phi)]$, where
\beqn
 J_0(z) & = & \delta(1+z)-\delta(1-z) \ , \nonumber \\
 J_{\pm m}(z,\phi) & = & (1-z^2)^{m/2}\delta^{(m)}(1\mp z)\cos[m(\phi-\phi_m)] \ .
 \label{mpole+}
\eeqn
Thus, we have $\Phi_{22}=0$ everywhere but on the singular null lines $u=0$, $z=\pm 1$, which are the histories of massless point particles generating the gravitational impulse. According to (\ref{mpole+}), these have a multipolar structure depending on $m$. In particular, the axially symmetric monopole term  $\,J_0$ represents a pair of particles with equal and opposite energy densities, localized at the two poles of the spherical wave front. 

On the other hand, the non-singular function
$  H(z,\phi)=b_0z+b_1\sqrt{1-z^2}\cos(\phi-\phi_1)$ satisfies
$\Psi_4=0$,  representing  thus  an impulse of null matter without a gravitational wave.

\subsubsection*{The $\Lambda<0$ case}

For $\>\epsilon_1=-1$, $\epsilon_2=0\,$ the relation (\ref{cond}) implies
$\frac{1}{2}a^{-2}=-\Lambda$ so that the metric (\ref{metric}) takes the form
\beqn
\d s^2=2\,\d\zeta\d\bar{\zeta}
+\frac{2\,\d u\d v - H(\zeta,\bar{\zeta})\,\delta(u)\,\d u^2}{(1+\Lambda uv)^2} \ .
\label{four-}
\eeqn
The Weyl and Ricci scalars which represent radiation are
$\Psi_4=\textstyle{\frac{1}{2}}H_{\zeta\zeta}\delta(u)$,
$\Phi_{22}=\textstyle{\frac{1}{2}}H_{\zeta\bar{\zeta}}\,\delta(u)$.
In this case the gravitational and/or pure radiation impulse propagates in the Pleba\'nski--Hacyan universe which is the direct product AdS$_2\times\mathbb{R}^2$ with isometries $SO(2,1)\times E(2)$ (the coordinate form of \cite{PlebHac79} is recovered after the transformation $w=v(1+\Lambda uv)^{-1}$). Using (\ref{cons}), the impulsive manifold  $U=0$ corresponds to $Z_2=\pm(-2\Lambda)^{-1/2}$. It is thus the history of {\em two non-intersecting 2-planes}.   Also, the embedding formalism makes it easy to verify that (\ref{four-}) admits the Killing vector $\pa_v+\Lambda u^2\pa_u$ for an arbitrary $H$.

The impulsive part of (\ref{four-}) describes a purely gravitational wave provided $\Phi_{22}=0$. The correspondence with  \pp waves enables us  to  use the results of \cite{GriPod97}, to which we refer for details. 
For a physical interpretation of the general solution $H(\zeta,\bar{\zeta})=f(\zeta)+\bar{f}(\bar{\zeta})$, we now introduce polar coordinates $\zeta=\rho\, e^{i\phi}$ on the planar wave front. In terms of these coordinates, we may write
\be
 H(\rho,\phi)=  a_0+b_0\ln\rho+\sum_{m=1}^\infty \Big(\,b_m\,\rho^{m}+b_{-m}\,\rho^{-m}\Big)\cos[m(\phi-\phi_m)] \ .
 \label{purewaves-}
\ee
The  constant $a_0$ is removable via the transformation $u\to u\,[1-{\textstyle\frac{1}{2}}a_0\Lambda u\Theta(u)]^{-1},$ $v\to v+{\textstyle\frac{1}{2}}a_0\Theta(u)$. The term linear in $\rho$ does not represent waves but still has an objective geometrical meaning (see the next section and \cite{PlebHac79}). The term proportional to $\rho^2$ is a  ``plane'' wave, corresponding to a constant $\Psi_4$. The powers  $\rho^m$ for $m>2$ result in unbounded curvature at infinity and for \pp waves have been investigated elsewhere \cite{PodVes99}. 
The remaining terms in (\ref{purewaves-}) are singular at $\rho=0$, and can be interpreted as gravitational waves generated by null particles with a multipole structure. 
Indeed, the source distribution $J(z,\phi)$ now defined as $\Phi_{22}\equiv\frac{\pi}{4}\,J(z,\phi)\,\delta(u)$  turns out to be
\be
 J(\rho,\phi)=b_0\delta(\rho)-\sum_{m=1}^\infty b_{-m}\frac{(-1)^m}{(m-1)!}\,\delta^{(m)}(\rho)\cos[m(\phi-\phi_m)] 
 \ .
\ee
Hence, $\Phi_{22}=0$ everywhere but on the singular null line $u=0=\rho$. The monopole term $\,\delta(\rho)\,$ describes a single point source at the origin of each impulsive plane, and is the analogue of the Aichelburg--Sexl null particle in the case of impulsive \pp waves \cite{AicSex71}.

Impulsive pure radiation without gravitational waves arises when $H(\rho)=a_2\rho^2$,
for which $\Psi_4=0$ and $\Phi_{22}=\frac{1}{2}a_2\delta(u)$.

\section{Some generalizations}

By applying  a formalism analogous to that used for the construction of impulsive waves in the (anti-)de Sitter \cite{HotTan93,PodGri98cqg}, (anti-)Nariai and Bertotti--Robinson \cite{Ortaggio01} universes, we have introduced impulses into the    Pleba\'nski--Hacyan spacetimes.  This completes the list of all possible non-expanding impulsive waves in the spacetimes which are the  direct product of two constant-curvature 2-spaces.
Both the impulsive solutions (\ref{four+}) for $\Lambda>0$ and (\ref{four-}) for $\Lambda<0$ reduce to  impulsive \pp waves for a vanishing cosmological constant. We have also demonstrated that pure gravitational waves are generated by point sources with an arbitrary multipole structure. Since the field equation is linear, solutions can be constructed which contain an arbitrary number of arbitrary multipole sources distributed arbitrarily over the impulsive surface.

Moreover,  there also exist more general impulsive waves in the ``truly'' exceptional spacetime of \cite{PlebHac79} with $\Lambda<0$, which is \emph{not} a direct product spacetime  (\ref{metric}). The corresponding metric is
\be
 \d s^2=2\,\d\zeta\,\d\bar\zeta+2\,\d u\,\d w+\left[2\Lambda w^2+\zeta L+\bar\zeta\bar L-H(\zeta,\bar\zeta)\delta(u)\right]\d u^2 \ ,
 \label{four-2}
\ee
where $L(u)$ is an arbitrary complex function. This reduces to  the solution (\ref{four-}) when $L=0$, after performing the transformation $w=v(1+\Lambda uv)^{-1}$. A non-vanishing  $L$ makes the second double null direction of the background non-geodesic (although still shear-free), and reduces the number of symmetries \cite{PlebHac79}. Nevertheless, it does not enter the curvature scalars calculated in the null tetrad ${\bf m}=\pa_{\bar{\zeta}}$, ${\bf l'}=-\pa_u+\frac{1}{2}\left[2\Lambda w^2+\zeta L+\bar\zeta\bar L-H\delta(u)\right]\pa_w$, ${\bf k'}=\pa_w$.
In particular, pure  gravitational waves are again given by (\ref{purewaves-}).

Finally, the above families of impulsive metrics (\ref{four+}), (\ref{four-}) and its generalization (\ref{four-2}) can be understood as distributional limits of exact Kundt waves with an \emph{arbitrary profile} function $H(\zeta,\bar\zeta,u)$. In particular, the expressions (\ref{purewaves+}) and (\ref{purewaves-}), which describe pure gravitational waves, remain valid. Note that such solutions without pure radiation, given by   $H=f(\zeta,u)+\bar f(\bar\zeta,u)$,  were already considered by Garc\'{\i}a and Alvarez \cite{GarAlvar84}. An interesting spacetime exists also in the pure radiation sub-family. This generalization of the impulsive solution (\ref{four-}) with $H(\rho)=a_2\rho^2$ is given by the metric
\be
 \d s^2=2\,\d\zeta\,\d\bar\zeta+2\,\d u\,\d w+\left[2\Lambda w^2+\zeta L+\bar\zeta\bar L-\zeta\bar\zeta\, d(u)\right]\d u^2 \ ,
 \label{planeem}
\ee
for which $\Lambda\le0$, $\Psi_4=0$, and $\Phi_{22}=\frac{1}{2}d(u)$. It thus describes a non-singular plane-fronted wave of null matter with an arbitrary profile  $d(u)$ which propagates in the non-trivial Pleba\'nski--Hacyan electrovac background. Interestingly, for $\Lambda=0$ this becomes a well-known conformally flat plane wave solution of the Einstein--Maxwell equations  \cite{kramerbook}.

\addvspace{1cm}

\section*{Acknowledgments}

The work was supported by  the grants GACR-202/02/0735
from the Czech Republic,  GAUK 141/2000 of Charles University in Prague, and by INFN.

\vspace{.5cm}


\end{document}